\documentclass[sn-aps, Numbered, iicol]{sn-jnl}

\usepackage{graphicx}%
\usepackage{multirow}%
\usepackage{tabularx}%
\usepackage{amsmath,amssymb,amsfonts}%
\usepackage{amsthm}%
\usepackage{mathrsfs}%
\usepackage[title]{appendix}%
\usepackage{xcolor}%
\usepackage{textcomp}%
\usepackage{manyfoot}%
\usepackage{booktabs}%
\usepackage{algorithm}%
\usepackage{algorithmicx}%
\usepackage{algpseudocode}%
\usepackage{listings}%

\theoremstyle{thmstyleone}%
\theoremstyle{thmstyletwo}%

\theoremstyle{thmstylethree}%

\begin{document}

\title{Quantum Convolutional Autoencoders for Reconstruction-Based Anomaly Detection}

\author*[1,3]{\fnm{Donovan} \sur{Slabbert}}\email{donovanslab@mweb.co.za}

\author[1,2,3]{\fnm{Francesco} \sur{Petruccione}}\email{petruccione@sun.ac.za}

\affil*[1]{\orgdiv{Department of Physics}, \orgname{Stellenbosch University}, \city{Stellenbosch}, \postcode{7600}, \country{South Africa}}

\affil[2]{\orgdiv{School of Data Science and Computational Thinking}, \orgname{Stellenbosch University}, \city{Stellenbosch}, \country{South Africa}}

\affil[3]{\orgname{National Institute of Theoretical and Computational Sciences (NITheCS)}, \city{Stellenbosch}, \country{South Africa}}

\abstract{Quantum convolutional neural networks (QCNNs) have become increasingly popular in quantum machine learning (QML) due to their efficient parameterization and hierarchical representation of quantum information. Anomaly detection is an important machine learning task with applications across a wide range of domains, including scientific data analysis. In this work, we adapt a QCNN architecture into a quantum autoencoder (QAE) framework for reconstruction-based anomaly detection. The models are trained in a semi-supervised manner on normal samples to reconstruct feature-extracted and dimensionally reduced time-series data, with reconstruction error used as an anomaly score. We investigate two quantum convolutional autoencoder architectures that differ in their treatment of latent information: a hierarchical architecture in which information remains distributed across the circuit and a bottleneck-based architecture in which information is explicitly compressed and reconstructed using additional decoder qubits. The size of the quantum latent space is varied to study its influence on reconstruction accuracy and anomaly detection performance. The approaches are benchmarked against both a variational quantum circuit and a comparable classical baseline using the Kepler exoplanet dataset and pulsar candidates from the High Time Resolution Universe Survey (HTRU-2 Dataset). The models are evaluated over ten deterministic optimization seeds at matched four- and eight-component input dimensions and matched latent-space sizes where feasible. Results indicate a trade-off between latent-space size and model capacity, while also suggesting that explicit latent-space compression through a quantum bottleneck can improve anomaly detection performance relative to architectures that retain information throughout the circuit. The strongest classical autoencoder performs better than both quantum autoencoder families on HTRU-2, while the supervised neural network achieves the strongest overall discrimination on both datasets.}




\maketitle

\section{Introduction}\label{intro}

Quantum machine learning (QML) is a field that combines the principles used in quantum computing with classical machine learning \cite{schuld2021machine, schuld2015introduction, biamonte2017quantum}. By exploiting quantum phenomena such as superposition and entanglement, QML aims to develop machine learning methods that may differ fundamentally from their classical counterparts. The hope is that these new methods could potentially lead to improvements in specific ways \cite{biamonte2017quantum, cerezo2022challenges, dunjko2018machine}, such as performance increases or runtime gains. These possibilities have led to the application of QML to various tasks such as classification \cite{senokosov2024quantum, abohashima2020classification, kavitha2024quantum}, clustering \cite{kerenidis2021quantum, lloyd2013quantum, kerenidis2019q, li2022quantum, slabbert2026quantum}, and even generative models \cite{lloyd2018quantum, tian2023recent, gao2018quantum, zoufal2021generative, dallaire2018quantum}. Among the most widely studied QML approaches are variational quantum machine learning methods and quantum kernel methods \cite{cerezo2021variational, havlivcek2019supervised, schuld2019quantum, huang2021power}. While these approaches have demonstrated promise for certain applications, their practical utility remains an active area of research. Challenges including trainability issues such as barren plateaus \cite{mcclean2018barren, cerezo2021cost}, sensitivity to noise on near-term quantum hardware \cite{bharti2022noisy, preskill2018quantum}, scalability limitations \cite{cerezo2022challenges}, and, in some cases, the possibility of efficient classical simulation \cite{bermejo2026quantum, shin2024dequantizing} have made it difficult to demonstrate consistent quantum advantages in practice. Nevertheless, QML remains an active research area and is widely explored as a framework for implementing machine learning models on current and emerging quantum devices.

Anomaly detection is a specific machine learning task in which the goal is to flag observations or samples that deviate significantly from the majority of the data \cite{chandola2009anomaly, nassif2021machine}. It is relevant for applications where certain events are rare or labels are limited, such as fraud detection \cite{ali2022financial} and industrial monitoring \cite{saufi2019challenges}. In the context of astronomy specifically, anomaly detection is particularly relevant to discovering rare or unexpected phenomena \cite{henrion2012classification}, such as unusual stars \cite{pashchenko2018machine}, galaxies \cite{gauci2010machine}, or transient events \cite{neira2020mantra}, with these examples spanning both anomaly detection and broader astronomical machine learning. Real-world astronomical datasets are often imbalanced \cite{baron2019machine}, making automated anomaly detection methods important for efficiently detecting anomalies within large-scale surveys. In our case, the tasks are exoplanet detection, where anomalous signals in stellar light curves can indicate the presence of previously unknown exoplanets \cite{malik2022exoplanet}, and pulsar candidate selection using data from the High Time Resolution Universe Survey (HTRU-2 Dataset), where true pulsars are treated as anomalies \cite{misc_htru2_372,lyon2016fifty}.

Quantum machine learning has been applied to anomaly detection before \cite{corli2025quantum, liu2018quantum}, where two specific examples include time series rewinding \cite{baker2025quantum}, where anomalies in sequential data are identified by reversing the learned Hamiltonian dynamics and detecting deviations from expected behaviour, and hybrid quantum-classical autoencoders \cite{sakhnenko2022hybrid}, which typically incorporate classical layers. Certain quantum autoencoders have been used in reconstruction-based frameworks, where the reconstruction error serves as an anomaly score for detecting outliers \cite{frehner2025applying, wang2022data, wu2024quantum}. Applications to large-scale, real-world scientific datasets remain limited. In particular, studies employing quantum convolutional autoencoders with progressively compressed latent representations for anomaly detection appear to be scarce in the current literature. This forms the basis of the approach investigated in this work, which explores quantum convolutional autoencoders for reconstruction-based anomaly detection, with particular emphasis on the role of latent-space compression and a comparison between architectures that retain information throughout the circuit and architectures that employ an explicit latent bottleneck.

We implement two variants of a quantum convolutional autoencoder architecture \cite{cong2019quantum} that operate on dimensionally reduced scientific data: features extracted from the Kepler time series and the numerical pulsar-candidate features supplied with HTRU-2 \cite{kepler_kaggle,misc_htru2_372,lyon2016fifty}. The models reconstruct the input feature vectors, and the resulting reconstruction error is used as an anomaly score to identify deviations from expected behaviour. The first employs a hierarchical quantum convolutional structure in which information is progressively concentrated onto a smaller subset of qubits while all input qubits remain physically present and are reactivated during decoding. The second follows a more conventional autoencoder design, where information is compressed into a bottleneck, after which the discarded input qubits are excluded from decoder operations and reconstruction is performed using additional decoder qubits. The primary goal is to determine how latent-space size and treatment affect reconstruction and anomaly-detection performance. The autoencoders are compared at matched input and latent dimensions and benchmarked against a variational quantum circuit and classical baselines on both datasets.

Differences in area under the receiver operating characteristic curve (ROC AUC) are additionally assessed through exact paired tests on matched optimization seeds, with Holm adjustment applied within each defined comparison family.

This paper is outlined as follows: Section \ref{sec:theory} covers the theory necessary for the implementation of the quantum convolutional autoencoders. Section \ref{sec:method} outlines all the methods used. Section \ref{sec:results} illustrates and discusses all results. Section \ref{sec:conc} is the conclusion.

\section{Theory}\label{sec:theory}

\subsection{Quantum Convolutional Neural Networks (QCNN)}

Quantum convolutional neural networks (QCNNs) \cite{cong2019quantum, hur2022quantum} are a sub-class of variational quantum circuits (VQCs) \cite{cerezo2021variational}. QCNNs have an inherent hierarchical structure, which arises from the progressive reduction of the number of qubits through alternating convolution and pooling layers. The physical register can remain allocated while later layers act on fewer qubits. The layers themselves are multi-qubit unitary operators, with architectures that can be designed or optimized.

The first step is the state preparation step, where the input feature vector is encoded into a quantum state. A classical feature vector $\mathbf{x} \in \mathbb{R}^d$ is embedded or encoded into an $n$-qubit state using an encoding unitary $U_\text{enc}(\mathbf{x})$:

\begin{equation}
    |\psi_0 \rangle = U_\text{enc} (\mathbf{x}) |0 \rangle^{\otimes n}.
\end{equation}

One common choice is angle encoding, where each feature is assigned to a single qubit through the rotation of that qubit. In other words, each qubit is rotated by a rotational parameter equal to the feature value. This type of encoding requires the number of features to match the number of qubits, which implies that $n = d$. The corresponding angle encoding unitary is:

\begin{equation}
    U_\text{angle} (\mathbf{x}) = \bigotimes_{i=1}^{n} R_X(x_i),
\end{equation}

where each feature $x_i$ is encoded as the rotation angle of an $R_X$ gate applied to qubit $i$. Amplitude encoding is also possible, but we will focus on angle embedding as this is the encoding used throughout this work.

Following state preparation, the encoded state is processed through $L$ layers of alternating convolutional and pooling layers. Each convolutional layer applies an ansatz that contains a combination or set of trainable unitaries $U_\text{conv}^{(l)}(\boldsymbol{\theta}^{(l)})$ to a specific set of qubits. The pooling operations, implemented using entangling gates such as controlled-NOT (CNOT) or controlled-rotation gates, reduce the number of qubits participating in subsequent encoder layers from $n_l$ to $n_{l+1}$ ($n_{l+1} < n_l$) at each layer. This dimensionality reduction introduces a type of coarse-graining of information, analogous to classical pooling in convolutional neural networks. 

The hierarchical arrangement has the consequence that some features of the quantum state undergo more gates in later layers, whereas others are affected less. As a result, certain qubits might capture finer details of the input, whereas others could represent broader correlations. This behavior is naturally tied to the concept of light cones in quantum machine learning \cite{suzuki2025light}. Light cones refer to the subset of input qubits and quantum operations that can influence a particular measurement outcome. The operation of one set of convolutional and pooling layers is therefore represented as:

\begin{equation}
    |\psi_{l}\rangle = U_\text{pool}^{(l)} U_\text{conv}^{(l)}(\boldsymbol{\theta}^{(l)}) |\psi_{l-1}\rangle, \quad l = 1, \dots, L.
\end{equation}

Convolutional layers can reuse the same trainable unitaries across multiple layers, with or without weight sharing. Weight sharing reduces the total number of independent parameters, making the network more resource-efficient, at the cost of model capacity. Additionally, because pooling reduces the number of active qubits, each subsequent layer requires fewer parameters, further reducing computational cost. This combination of weight sharing and progressive qubit reduction gives rise to a hierarchical quantum feature representation. In other words, information captured at early layers influences later layers, and increasingly abstract features are encoded in the smaller latent system. Different qubits therefore capture abstract information at varying levels of specificity, with some encoding highly specialized features and others maintaining more general representations.

At the final layer, the remaining qubits are measured to produce outputs, which can correspond to classification logits, probabilities, or other observables depending on the task. Typically, the $Z$-expectation values of the qubits are measured. The variational parameters $\boldsymbol{\theta} = \{\boldsymbol{\theta}^{(1)}, \dots, \boldsymbol{\theta}^{(L)}\}$ are optimized using a classical optimizer in a hybrid quantum-classical training loop, minimizing a task-specific loss function such as cross-entropy for classification or mean squared error for reconstruction tasks.

\subsection{Quantum Convolutional Autoencoders (QAE)}

The quantum convolutional autoencoders (QAEs) considered extend the QCNN framework above to perform feature vector reconstruction rather than classification. After a number of convolution and pooling layers, a subset $k \leq n$ of qubits defines a latent quantum state $|\psi_\text{lat}\rangle \in \mathcal{H}_{2^k}$, which forms a compressed representation for $k<n$ or a full-latent representation for $k=n$ of the input. The latent space qubits capture the essential information required to reconstruct the original feature vector. The QCNN that implements this is called the encoder and is represented as:

\begin{equation}
    |\psi_\text{lat}\rangle = \prod_{l=1}^{L} U_\text{enc}^{(l)}(\boldsymbol{\theta}^{(l)}) |\psi_0\rangle.
\end{equation}

Here $|\psi_0\rangle \in \mathcal{H}_{2^n}$ is the encoded input state and $U_\text{enc}^{(l)}(\boldsymbol{\theta}^{(l)})$ denotes the parametrised encoder unitary at layer $l$. Instead of measuring the latent qubits for a classification task, the QAE appends variational decoder layers $U_\text{dec}^{(l)}(\boldsymbol{\phi}^{(l)})$, which progressively expand the latent space back to $n$ reconstruction outputs on persistent or additional decoder qubits:

\begin{equation}
    |\hat{\psi}_0\rangle = \prod_{l=1}^{L} U_\text{dec}^{(l)}(\boldsymbol{\phi}^{(l)}) |\psi_\text{lat}\rangle.
\end{equation}

Here $U_\text{dec}^{(l)}(\boldsymbol{\phi}^{(l)})$ denotes the parametrised decoder unitary at layer $l$, acting on the latent state to reconstruct the original system. For each reconstructed component, the measured Pauli-$Z$ expectation value $z_i=\langle \hat{Z}_i \rangle$ lies in $[-1,1]$. Since the input target is an angle $\theta_i\in[0,\pi]$, the corresponding reconstructed angle is

\begin{equation}
    \widehat{\theta}_i = \arccos\!\left[\operatorname{clip}\!\left(z_i,-1+\epsilon,1-\epsilon\right)\right], \qquad \epsilon=10^{-7}.
\end{equation}

Clipping prevents floating-point values marginally outside $[-1,1]$ from making the inverse cosine undefined. The reconstruction loss is then evaluated directly in angle space:

\begin{equation}
    \mathcal{L}_\text{rec} = \frac{1}{d} \sum_{i=1}^{d} \left(\theta_i - \widehat{\theta}_i \right)^2.
\end{equation}

The mapping $z_i=\cos(\theta_i)$ is monotonic on $[0,\pi]$, so the $Z$-expectation value uniquely determines the encoded angle on the fitted domain.

This reconstruction error can then be used as an anomaly score, where higher errors indicate that a sample deviates more from the learned normal data distribution. The latent-space size ($k$) influences the capacity of the autoencoder by controlling the amount of information that can be retained within the compressed representation. Smaller latent spaces impose stronger compression and encourage the model to preserve only the most important features of the input, whereas larger latent spaces allow more detailed information to be retained. Consequently, varying the number of latent qubits provides a means of studying the trade-off between compression and reconstruction fidelity.

The role of the latent space differs between the two QAE variants considered in this work. In the hierarchical QCNN-based architecture, information is progressively concentrated onto a smaller subset of qubits, but the remaining qubits continue to persist throughout the circuit. Although these qubits do not form part of the latent representation, they can still influence the reconstruction process through the network's hierarchical structure and associated light cones. Consequently, reconstruction may depend not only on the latent representation itself, but also on correlations preserved elsewhere in the circuit. In contrast, the bottleneck-based architecture employs a strict latent representation, where the latent qubits form the sole compressed representation used for reconstruction. Information discarded during compression is therefore generally irrecoverable, making reconstruction dependent entirely on the information retained within the latent space.

Both approaches leverage the hierarchical structure of the QCNN to learn compact representations of the input while reconstructing the original feature vector through a variational decoder. The bottleneck architecture achieves this through explicit compression into a reduced latent space, whereas the hierarchical architecture retains additional information pathways through persistent qubits. In both cases, reconstruction error provides a natural mechanism for anomaly detection, with anomalous samples expected to reconstruct less accurately than samples drawn from the learned normal distribution. The combination of convolutional unitaries, pooling operations for dimensionality reduction, and variational decoding therefore forms the basis of the QAE architectures considered in this work.

\section{Methodology}\label{sec:method}

\subsection{Feature Extraction and Preprocessing}

The original Kepler dataset \cite{kepler_kaggle} consists of labeled time-series flux measurements from the Kepler Space Telescope. The labels correspond to the presence (label 1, anomaly) or absence (label 0, normal) of exoplanet transits. Each sample contains 3197 sequential flux measurements representing stellar brightness over time. The dataset is provided as separate training and test sets with shapes $(5087,3198)$ and $(570,3198)$, respectively, including the label column. The supplied split was retained and contains 5050 normal samples and 37 anomalies in the training set, and 565 normal samples and five anomalies in the test set. Prior to feature extraction, each light curve was independently rescaled to the range $[-1,1]$ using sample-wise min-max normalization, and the original labels were relabeled from ${1,2}$ to ${0,1}$. Labels were tracked throughout all preprocessing stages to ensure correct sample-to-label alignment for downstream anomaly-detection and classification tasks.

A second evaluation used the HTRU-2 pulsar-candidate dataset. It contains 17,898 observations described by eight numerical features, with 16,259 normal candidates and 1,639 pulsars treated as anomalies. A single stratified 80:20 partition with random state 1 was used for every model and optimization seed. The resulting training set contains 13,007 normal samples and 1,311 anomalies, while the test set contains 3252 normal samples and 328 anomalies.

Feature extraction was then performed on the Kepler time-series training and test sets using the \texttt{tsfresh} package \cite{christ2018time}, calculating a wide range of time-series features such as statistical moments (mean, variance, skewness, kurtosis), autocorrelation coefficients, linear trend parameters, and frequency-domain features. This resulted in a total of 777 features that captured the temporal patterns and statistical properties of the time series. Extracted features with constant or near-constant values were removed, and the remaining features were filtered based on their relevance to the training labels using \texttt{tsfresh} via a univariate relevance scoring procedure \cite{guyon2003introduction}, which retained only features most predictive of the target. This resulted in 211 features remaining. The resulting shapes were $(5087, 211)$ for the training set and $(570, 211)$ for the test set, excluding labels. The training labels were used only during feature selection to identify the most informative features. The retained Kepler feature set was fixed across all optimization seeds. HTRU-2 already provides eight derived numerical features, so these feature-extraction and relevance-selection steps were omitted for that dataset.

The remaining features were then standardized to have zero mean and unit variance. This normalized the contribution of each feature and reduced any bias due to differing magnitudes. Standardization was applied before principal component analysis (PCA) so that features on different numerical scales contributed comparably to the component directions. PCA was subsequently fitted to the standardized training features and applied to the test set to obtain lower-dimensional feature representations suitable for quantum encoding. Separate four- and eight-component representations were used for matched comparisons across the persistent quantum autoencoder (pQAE), bottleneck quantum autoencoder (bQAE), classical autoencoder (CAE), VQC, and neural network (NN) model families. The resulting PCA components were then scaled to the range $[0,\pi]$ to match the input requirements of the quantum machine learning models. The standard scaler, PCA transformation, and component-wise min--max scaler were fitted using the complete training set and subsequently applied unchanged to the test set. For anomaly-detection experiments based on feature-vector reconstruction, only normal training samples (label 0) were used during autoencoder training, while the test set retained all samples to allow evaluation of anomaly-detection performance. For supervised methods, all training samples were used throughout the training pipeline.

The final preprocessing step mapped each retained component to the rotation-angle domain. For sample $i$ and component $j$, let $u_{ij}$ denote the PCA score and let $m_j$ and $M_j$ denote the minimum and maximum scores observed for that component in the complete training partition. The component-wise transformation was
\begin{equation}
    \theta_{ij}=\pi\frac{u_{ij}-m_j}{M_j-m_j}.
\end{equation}
The fitted transformation was then applied unchanged to the test scores. Test scores below $m_j$ or above $M_j$ were recorded before their transformed angles were clipped to $[0,\pi]$; angles were never wrapped. The Kepler test set contained no out-of-range samples. For HTRU-2, two test samples were affected at both input dimensions: one normal sample and one anomalous sample each contained one out-of-range component. The affected anomalous sample represents one of the 328 test anomalies, or $0.305\%$.

\subsection{Quantum approaches}

The main methods of this work are two types of quantum convolutional autoencoders based on previously proposed quantum autoencoders \cite{romero2017quantum, wu2024quantum, frehner2025applying, sakhnenko2022hybrid} and incorporating quantum convolutional neural network (QCNN) structures \cite{cong2019quantum}. The first type, referred to as persistent quantum autoencoders (pQAEs), employs a hierarchical QCNN architecture in which all qubits remain present throughout the circuit while information is progressively concentrated onto a smaller subset of latent qubits. The second variant, referred to as bottleneck quantum autoencoders (bQAEs), employs a conventional latent bottleneck in which reconstruction is performed from a compressed latent representation using additional decoder qubits. The pQAE was evaluated in seven input-latent configurations and the bQAE in six configurations, with matched input and latent dimensions used wherever feasible.

The objective of all autoencoder models was feature-vector reconstruction, where the reconstruction error served as both the training loss and the anomaly score used for anomaly detection. Reconstruction error was quantified using the mean-squared error (MSE) between the input and reconstructed feature vectors. All models operated on feature vectors derived from principal component analysis (PCA). Both QAE families were evaluated using four-component inputs with latent sizes $k=4,2,1$. Eight-component inputs used $k=8,4,2,1$ for the pQAE and $k=8,4,2$ for the bQAE. This design choice was motivated by the increased qubit requirements of the bottleneck architecture, which introduces additional qubits during decoding and would therefore require substantially more qubits to accommodate an eight-component input representation. The bQAE with eight input components and one latent qubit requires 15 physical qubits. Although feasible in principle, this configuration was omitted from the reported experiments because its simulation time and memory requirements became impractical. A bQAE with $n$ inputs and $k$ latent qubits uses $2n-k$ physical qubits, while the pQAE uses $n$. In the bQAE, discarded encoder qubits remain allocated in the simulated register but are excluded from decoder operations. The $n$ reconstruction outputs comprise the $k$ latent qubits and $n-k$ additional decoder qubits initialized in $|0\rangle$.

The encoding strategy prior to any convolution or pooling operations was angle encoding using $R_X$ rotations, where the rotation angles were given by the PCA component values scaled to the range $[0,\pi]$. The convolutional layers use trainable $R_X$ rotations, pooling uses CNOT operations, and a single trainable $R_Y$ layer is applied to the latent qubits at the encoder--decoder boundary. Each layer and active width uses an independent parameter set. The ordering of encoded features may influence the information retained near the latent space due to the hierarchical reduction structure of the architectures. Consequently, PCA components with higher explained variance were preferentially assigned to qubits that persist longer through the convolution and pooling operations, such that more informative features remain closer to the latent representation. While this design choice is motivated by the hierarchical nature of the architectures, its optimality is not formally established and should be regarded as a heuristic consideration rather than a strict requirement. The PCA columns were reversed so that the first principal component was assigned to input position $n-1$, the second to $n-2$, and so forth. The first four and eight components explain $64.35\%$ and $75.76\%$ of the Kepler training variance, and $94.29\%$ and $100.00\%$ of the HTRU-2 training variance, respectively.

\begin{figure*}[ht!]
    \centering
    \includegraphics[width=\textwidth]{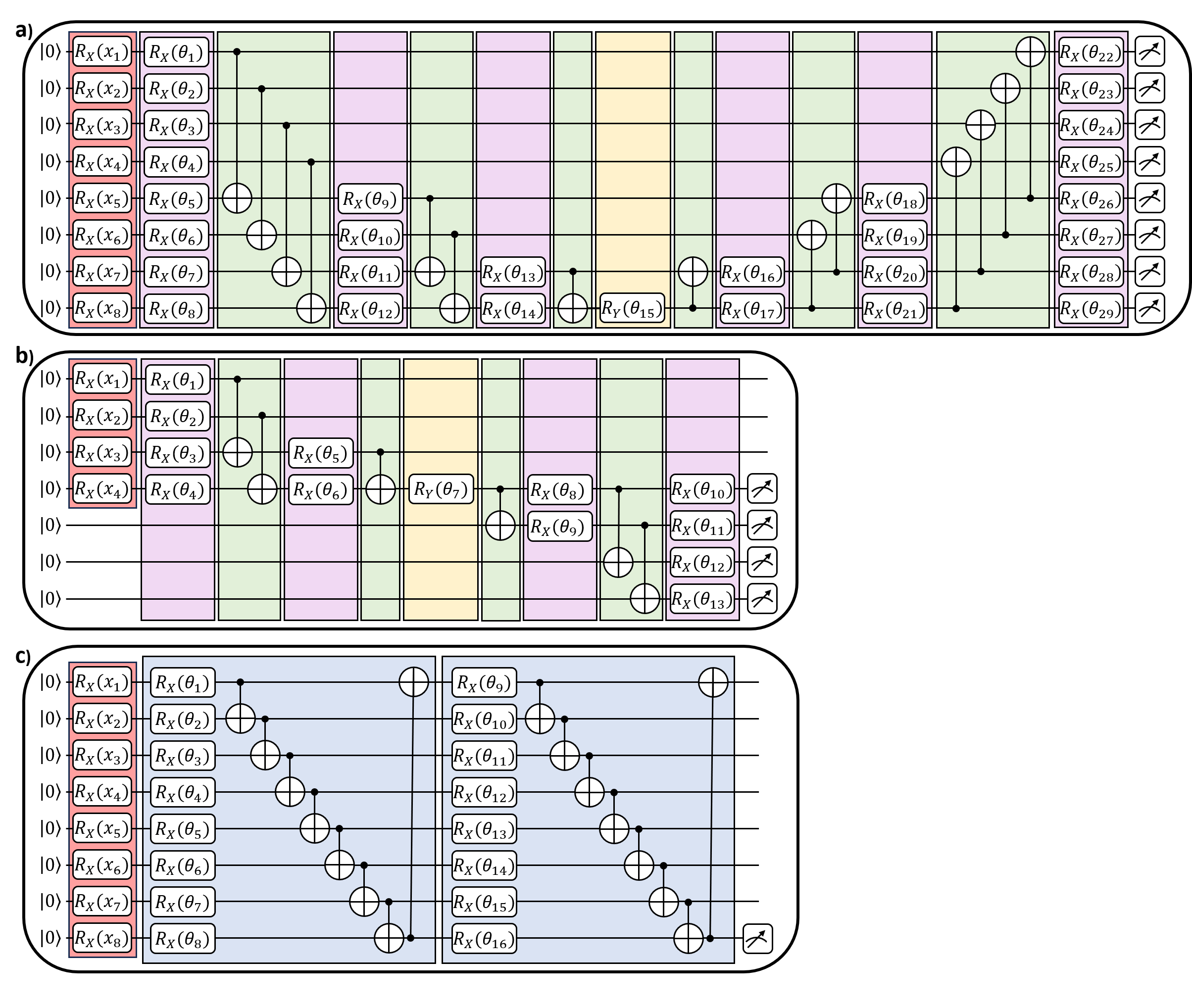}
    \caption{Example quantum circuits used in this study: (a) an eight-input pQAE with one latent qubit, (b) a four-input bQAE with one latent qubit, and (c) an eight-input VQC with two basic entangling layers. The displayed circuits are representative examples of the evaluated model families. Red denotes angle encoding, pink denotes trainable rotational gates forming the convolutional operations, green denotes pooling operations, yellow denotes the latent-space operation, and blue denotes a basic entangling layer. In the QAE circuits, the Pauli-$Z$ expectation values of all designated output qubits are measured and inverse-cosine decoded before computing the MSE reconstruction loss. In the VQC circuit, only the final qubit is measured, with $P(1)$ used as the anomaly-class probability in the BCE loss function.}
    \label{fig:concept}
\end{figure*}

Following the decoding stage, reconstruction outputs were obtained from the $Z$-expectation values of the designated output qubits. These expectation values were subsequently clipped to $[-1+\epsilon,1-\epsilon]$ and converted to reconstructed angles using the inverse cosine to match the feature scaling used during encoding. The convolutional layers employed trainable $R_X$ rotations, while pooling was implemented using CNOT-based entangling operations that progressively reduced the number of active qubits. The specific encoder and decoder structures differed between the pQAE and bQAE architectures and are summarized in Table~\ref{table:resources}. A conceptual illustration of the architectures with a single latent qubit is shown in Figure~\ref{fig:concept}.

The distinct autoencoder architectures differ primarily in the size of the latent space and the corresponding number of compression stages required to reach that representation. The number of compression stages depends on both the number of encoded input qubits and the desired latent-space dimensionality. Consequently, architectures with the same latent-space size do not necessarily have the same circuit depth. For example, reducing eight input qubits to one latent qubit requires three convolution-and-pooling stages, whereas reducing four inputs to one requires two. When $k=n$, convolution and pooling are omitted; these full-latent configurations retain the input encoding and trainable latent rotations and therefore serve as no-compression control architectures rather than identity maps. In all cases, there exists an implicit trade-off between latent-space size, circuit depth, and information retention. Smaller latent spaces provide stronger compression but may discard useful information, whereas larger latent spaces preserve more information at the expense of reduced compression. Deeper circuits may also increase expressibility, although they are generally more susceptible to noise and optimization difficulties.

An additional quantum method was implemented as a benchmark model. Separate four- and eight-component VQCs were evaluated. The benchmark consisted of a variational quantum circuit (VQC) \cite{benedetti2019parameterized} comprising two entangling layers followed by measurement of the final qubit. Each entangling layer consisted of a set of X-rotations on all qubits, each with an independent trainable parameter, and entangling gates or CNOT gates in a circular arrangement. This is similar to the QAE architecture with the circular arrangement of CNOTs. Two such layers were sequentially stacked, each with distinct parameter sets, to add to the expressive capacity of the model. The final qubit is measured, and the probability $P(1)$ is used directly as the anomaly-class probability in the weighted binary cross-entropy (BCE) loss. The quantum circuit for this quantum benchmark method is also illustrated in Figure~\ref{fig:concept}.

\begin{table}[t]
    \centering
    \setlength{\tabcolsep}{2pt}
    \fontsize{5pt}{7pt}\selectfont
    \caption{Resource counts for the model configurations considered. For quantum models, width is the number of physical qubits, depth follows the logical circuit-block convention used in the implementation, and the CNOT column gives the number of entangling gates. For classical models, width is the input-node count and depth is the number of linear layers. The 15-qubit bQAE $(8\rightarrow1)$ is shown for completeness and was excluded uniformly from the reported analysis.}
    \begin{tabular}{llrrrrr}
        \hline\hline
        Method & Input $n$ & Latent $k$ & Width & Parameters & CNOTs & Depth/layers \\
        \hline
        \multirow{7}{*}{pQAE} & 4 & 4 & 4 & 4 & 0 & 2 \\
         & 4 & 2 & 4 & 10 & 4 & 6 \\
         & 4 & 1 & 4 & 13 & 6 & 10 \\
         & 8 & 8 & 8 & 8 & 0 & 2 \\
         & 8 & 4 & 8 & 20 & 8 & 6 \\
         & 8 & 2 & 8 & 26 & 12 & 10 \\
         & 8 & 1 & 8 & 29 & 14 & 14 \\
        \hline
        \multirow{7}{*}{bQAE} & 4 & 4 & 4 & 4 & 0 & 2 \\
         & 4 & 2 & 6 & 10 & 4 & 6 \\
         & 4 & 1 & 7 & 13 & 6 & 10 \\
         & 8 & 8 & 8 & 8 & 0 & 2 \\
         & 8 & 4 & 12 & 20 & 8 & 6 \\
         & 8 & 2 & 14 & 26 & 12 & 10 \\
         & 8 & 1 & 15 & 29 & 14 & 14 \\
        \hline
        \multirow{2}{*}{VQC} & 4 & -- & 4 & 8 & 8 & 5 \\
         & 8 & -- & 8 & 16 & 16 & 5 \\
        \hline
        \multirow{7}{*}{CAE} & 4 & 4 & 4 & 40 & -- & 2 \\
         & 4 & 2 & 4 & 62 & -- & 4 \\
         & 4 & 1 & 4 & 69 & -- & 6 \\
         & 8 & 8 & 8 & 144 & -- & 2 \\
         & 8 & 4 & 8 & 220 & -- & 4 \\
         & 8 & 2 & 8 & 242 & -- & 6 \\
         & 8 & 1 & 8 & 249 & -- & 8 \\
        \hline
        \multirow{2}{*}{NN} & 4 & -- & 4 & 45 & -- & 3 \\
         & 8 & -- & 8 & 153 & -- & 3 \\
        \hline\hline
    \end{tabular}
    \label{table:resources}
\end{table}

\subsection{Classical approaches}

Classical analogues of the quantum methods were also implemented. It was important to design the classical model architectures to be as representative of the quantum models as possible. A classical autoencoder (CAE) was implemented using fully connected layers and a symmetric encoder-decoder structure. The architecture employed an explicit latent bottleneck analogous to the bQAE models, while the progression of layer widths was chosen to mirror the latent-space sizes investigated in the pQAE architectures. Four-component inputs used latent-space sizes of 4, 2, and 1 nodes, while eight-component inputs used sizes of 8, 4, 2, and 1 nodes.

The four- or eight-dimensional PCA feature vector was progressively reduced in a manner analogous to the quantum architectures. The encoder maps the input to a lower-dimensional latent representation using fully connected linear layers with rectified linear unit (ReLU) activations, and the decoder reconstructs the original feature vector from the latent space.

The smaller the latent space, the more fully connected layers were required to reach the desired compression, mirroring the increased number of compression stages present in the corresponding quantum architectures. Consequently, there exists a trade-off between model capacity, network depth, and information retention. Smaller latent spaces provide stronger compression but may discard useful information, whereas larger latent spaces preserve more information at the expense of reduced compression. The reconstructed outputs were compared directly with the input angle vectors in $[0,\pi]$ in the reconstruction loss function. The final decoder layer was linear and its output remained unconstrained.

A classical neural network (NN) was implemented as a benchmark classification model analogous to the variational quantum circuit benchmark. The NN consisted of two fully connected hidden layers with a width matching the four- or eight-component input and ReLU activations. A scalar logit was used as its output. The model was trained using supervised label information and binary cross-entropy (BCE) loss. Both the CAE and NN operated directly on the PCA-reduced feature vectors, allowing a direct comparison with the quantum approaches.

\subsection{Training and Evaluation}

All models, quantum and classical, were trained on PCA-reduced feature representations. Every model family was evaluated with four- and eight-component representations, subject to the excluded 15-qubit bQAE described above. PCA was fitted using the complete training set and subsequently applied to the test set. For reconstruction-based anomaly-detection methods, only normal training samples (label 0) were used during model training, whereas supervised classification methods were trained using all available training samples.

Training was performed for 100 epochs using the Adam optimizer with a learning rate of 0.001 and a batch size of 64, unless stated otherwise. Training samples were shuffled deterministically at each epoch. The epoch loss was recorded as the sample-weighted mean of all minibatch losses, and the final-epoch parameters were used for evaluation. For purely supervised methods, the class imbalance was mitigated by using a positive class weighting term $w = \frac{N_{\text{neg}}}{N_{\text{pos}}}$ that was incorporated into the loss-based objectives. Autoencoders used mean squared error (MSE) as the reconstruction loss, while classifiers used weighted binary cross-entropy loss. Ten runs used deterministic model seeds 1000--1009 while the supplied Kepler split, the fixed HTRU-2 split, and every fitted preprocessing transformation were held constant. The quantum circuits were evaluated with \texttt{shots=None} using the noiseless \texttt{lightning.qubit} simulator. The reported seed variation therefore measures optimization variability on fixed test samples.

Evaluation of model performance was done consistently across all methods. For the autoencoders, the mean squared reconstruction error served as the anomaly score of each sample. To investigate the sensitivity of reconstruction-based anomaly detection to the anomaly threshold, decision boundaries corresponding to the 99th, 95th, 80th, and 70th percentiles of the normal-training reconstruction-error distribution were evaluated. Samples with reconstruction errors above the selected percentile threshold were classified as anomalies. In addition to threshold-based evaluation, accuracy, precision, recall, and F1-score were computed for the autoencoders to allow direct comparison with the classification models. For all other models, predicted class probabilities were used to compute these metrics. Receiver operating characteristic (ROC) and precision--recall (PR) curves were generated for every model, and the corresponding area under the curve values (ROC AUC and PR AUC) were calculated. Average precision (AP) was also reported as a supplementary metric. ROC AUC, PR AUC, and average precision were calculated independently for every run from its raw labels and continuous scores. The plotted ROC and PR curves show the mean across the ten optimization seeds, with shaded bands indicating one standard deviation. Classification reports and confusion matrices were produced to connect the overall performance to the model predictions. Finally, training behavior was visualized using loss convergence curves. All of these results can be found in Section \ref{sec:results} in Figures~\ref{fig:losses},~\ref{fig:ROC_curves},~\ref{fig:PR_curves},~\ref{fig:ROC_PR}, and~\ref{fig:cm} and Tables~\ref{table:ad_metrics},~\ref{table:baseline_metrics},~\ref{table:matched_auc_tests}, and~\ref{table:selected_auc_tests}.

Pairwise differences in ROC AUC were evaluated using an exact two-sided paired sign-flip test. Runs were paired by deterministic optimization seed, and the test statistic was the mean of the ten within-seed differences, $d_r=\mathrm{AUC}_{A,r}-\mathrm{AUC}_{B,r}$. Assuming sign-exchangeability of the paired ROC AUC differences under the null hypothesis, all $2^{10}=1024$ possible sign assignments were enumerated to obtain exact two-sided $p$-values. These tests quantify the consistency of ROC AUC differences across optimization seeds on the same fixed test samples. Uncertainty from alternative data partitions and test-sample resampling remains outside their scope.

For the structurally matched comparisons, the family-wise error rate was controlled at $\alpha=0.05$ within each dataset-specific comparison family using the Holm procedure \cite{holm1979simple}. For each dataset, adjustment was applied separately within three structurally matched comparison families: four compressed bQAE--pQAE comparisons, six bQAE--CAE comparisons, and seven pQAE--CAE comparisons. These families address distinct scientific comparisons: the explicit-bottleneck and persistent quantum architectures are compared with one another, and each quantum architecture is compared separately with the classical autoencoder. The no-compression pQAE--bQAE cases were excluded from their direct family because the generated circuits are structurally identical when the latent size equals the input size. The unavailable bQAE $8\rightarrow1$ configuration was also excluded. For each dataset, a separate six-comparison Holm family compared the highest-mean-ROC-AUC pQAE, bQAE, and CAE with the VQC and NN at the same input dimension. These benchmark comparisons are data-selected because the same ten runs were used to choose the autoencoder configuration and perform the paired test. The Holm adjustment addresses multiplicity among the six reported tests but does not account for the preceding architecture-selection step. Accordingly, no study-wide family-wise error control is claimed.

Formal pairwise testing was restricted to ROC AUC because it is the primary threshold-independent metric used to rank the model configurations in this study. PR AUC and average precision are reported descriptively as complementary measures for the imbalanced datasets, while precision, recall, and F1 score additionally depend on the selected reconstruction threshold. Applying formal tests to every reported metric would substantially expand the multiple-comparison families. This concern is especially relevant for Kepler, where the threshold-based metrics are coarse because the test set contains only five anomalies.

\section{Results}\label{sec:results}

\subsection{Loss curves}

\begin{figure*}[ht!]
    \centering
    \includegraphics[width=\textwidth]{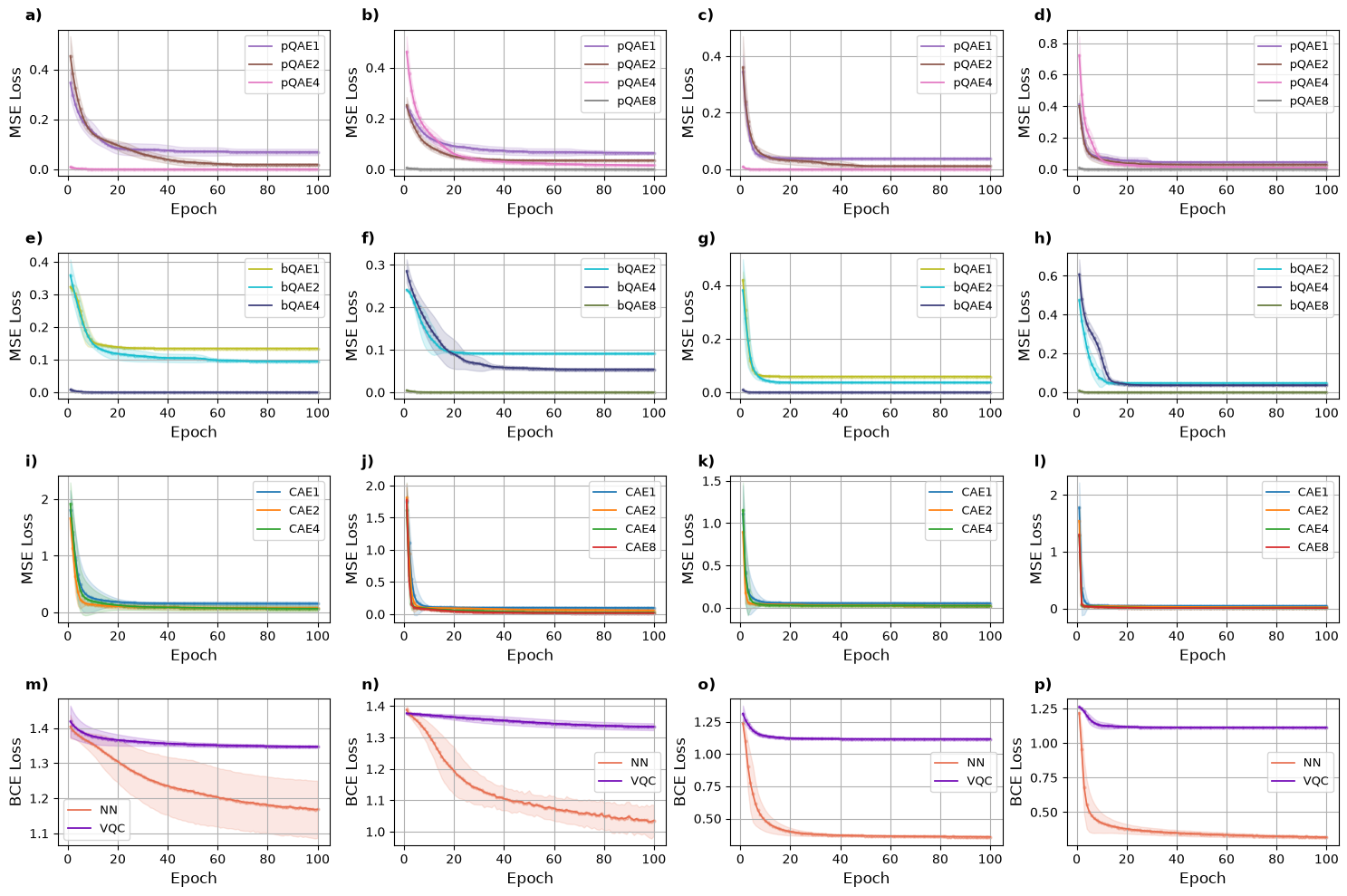}
    \caption{Mean training loss across ten optimization seeds over 100 epochs for all classical and quantum methods. Shaded regions indicate one standard deviation. Panels (a)--(d), (e)--(h), (i)--(l), and (m)--(p) show the pQAE, bQAE, CAE, and supervised NN/VQC models, respectively. Within every row, the columns show the Kepler four-component, Kepler eight-component, HTRU-2 four-component, and HTRU-2 eight-component experiments. Each autoencoder panel includes the available latent dimensions shown in its legend; the bQAE $8\rightarrow1$ configuration is omitted. The quantum and classical autoencoder panels report mean squared reconstruction error in angle space, while the supervised panels report weighted binary cross-entropy. The mean and standard deviation are calculated independently at every epoch across the ten fixed-split optimization-seed runs.}
    \label{fig:losses}
\end{figure*}

The loss curves shown in Figure~\ref{fig:losses} represent the mean training loss across ten optimization seeds, with shaded regions indicating one standard deviation. Mean squared error (MSE) loss is shown for the autoencoder models, while binary cross-entropy (BCE) loss is shown for the classification models. Absolute loss curves are shown for both Kepler and HTRU-2, allowing the convergence scale and seed-to-seed variability to be assessed directly for each configuration.

Overall, all models exhibit smooth convergence throughout training, although the degree of variability between runs differs across architectures. The uncertainty bands remain relatively narrow for most models, while larger fluctuations are observed for certain architectures, particularly for some of the HTRU-2 autoencoder configurations. For all autoencoder architectures, the majority of the loss reduction occurs during the early epochs of training, after which the optimization gradually approaches a plateau.

For the pQAE architectures, the loss curves in Figures~\ref{fig:losses}(a)--(d) show clear convergence for all latent-space sizes. The full-latent configurations approach reconstruction losses close to zero, while the compressed configurations approach higher stable plateaus. All pQAE architectures show that most of the reduction occurs during the earlier epochs.

The bQAE architectures likewise exhibit smooth and consistent convergence in Figures~\ref{fig:losses}(e)--(h). The full-latent controls reach losses close to zero. On Kepler, the mean final loss increases from 0.095 for $4\rightarrow2$ to 0.135 for $4\rightarrow1$, and from 0.054 for $8\rightarrow4$ to 0.091 for $8\rightarrow2$. On HTRU-2, the corresponding values increase from 0.037 to 0.059 for four-component inputs and from 0.036 to 0.048 for eight-component inputs. Stronger compression therefore produces higher final reconstruction loss across the available bQAE configurations. Most of the loss reduction occurs during the early epochs, and seed-to-seed variation remains small.

Similar behaviour is observed for the CAE architectures shown in Figures~\ref{fig:losses}(i)--(l), which exhibit rapid convergence during the initial stages of training. On Kepler, the mean final loss decreases from 0.153 to 0.074 to 0.065 across the $4\rightarrow1$, $4\rightarrow2$, and $4\rightarrow4$ configurations, and from 0.098 to 0.057 to 0.026 to 0.019 across the $8\rightarrow1$, $8\rightarrow2$, $8\rightarrow4$, and $8\rightarrow8$ configurations. HTRU-2 shows the same overall relationship between compression and reconstruction loss, with final means of 0.053, 0.024, and 0.023 for the four-component models and 0.049, 0.023, 0.007, and 0.008 for the eight-component models. The HTRU-2 $8\rightarrow4$ CAE finishes slightly below the full-latent $8\rightarrow8$ control. Several CAE configurations have wider uncertainty bands than the quantum autoencoders, especially on Kepler, indicating greater optimization-seed variability.

For the classification models shown in Figures~\ref{fig:losses}(m)--(p), both the VQC and NN exhibit relatively stable optimization behaviour throughout training. The VQC exhibits a particularly smooth optimization trajectory with relatively small uncertainty. Nevertheless, both models converge to stable loss values by the end of training. Since loss convergence alone does not determine predictive performance, a more detailed comparison based on evaluation metrics is presented in the following sections.

An interesting observation across the autoencoder families is that lower final reconstruction loss does not consistently correspond to stronger anomaly-detection performance. In particular, the full-latent controls reconstruct their inputs accurately while producing ROC-AUC values close to random discrimination. These results suggest that reconstruction quality alone is not a reliable predictor of downstream anomaly-detection capability. Consequently, evaluation based solely on reconstruction loss may not accurately reflect anomaly-detection performance.

\subsection{Receiver Operating Characteristic (ROC) and Precision–Recall (PR) Curves}

\begin{figure*}[ht!]
    \centering
    \includegraphics[width=\textwidth]{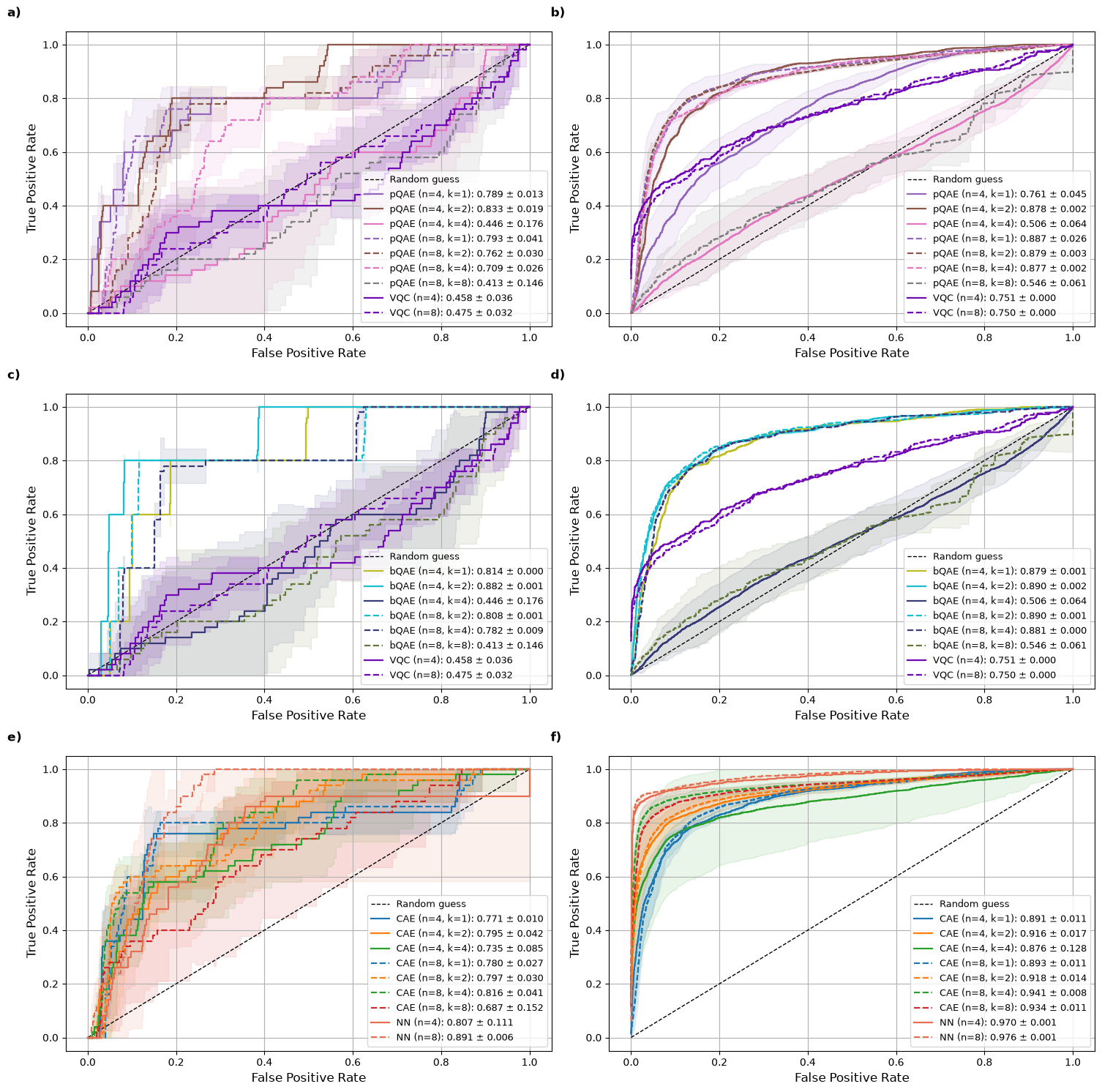}
    \caption{Mean ROC curves and corresponding ROC AUC values for all quantum and classical methods across ten optimization seeds. Shaded regions indicate one standard deviation. The random-guess classifier is indicated by the black dashed diagonal line, corresponding to a ROC AUC value of 0.5. The left column shows Kepler and the right column shows HTRU-2. The pQAE, bQAE, and CAE architectures are shown in (a,b), (c,d), and (e,f), respectively. The VQC is included for comparison in (a)--(d), while the NN is included in (e,f). Four- and eight-component inputs are distinguished by line style. For each optimization seed, the empirical ROC curve and ROC AUC were calculated directly from that run's raw test labels and continuous anomaly scores or anomaly-class probabilities. The displayed curve is the mean across seeds, while each reported ROC AUC is the mean $\pm$ standard deviation of the ten per-seed ROC AUC values and is not calculated from the displayed mean curve.}
    \label{fig:ROC_curves}
\end{figure*}

\begin{figure*}[ht!]
    \centering
    \includegraphics[width=\textwidth]{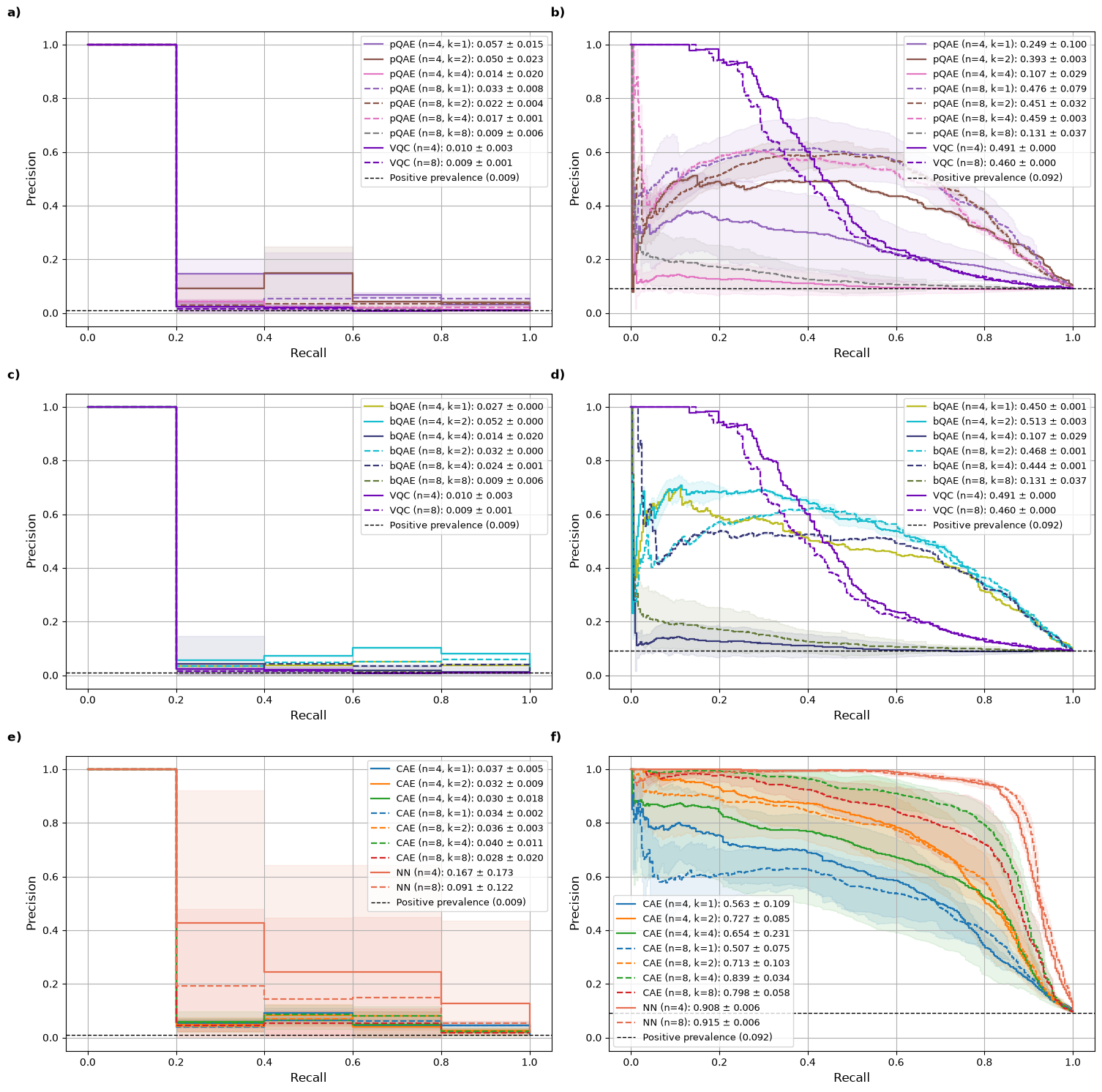}
    \caption{Mean precision--recall curves and corresponding PR AUC values for all quantum and classical methods across ten optimization seeds. Shaded regions indicate one standard deviation. The positive-class prevalence is indicated by the black dashed horizontal line. The left column shows Kepler and the right column shows HTRU-2. The pQAE, bQAE, and CAE architectures are shown in (a,b), (c,d), and (e,f), respectively. The VQC is included for comparison in (a)--(d), while the NN is included in (e,f). Four- and eight-component inputs are distinguished by line style. For each optimization seed, the empirical precision--recall curve and PR AUC were calculated directly from that run's raw test labels and continuous anomaly scores or anomaly-class probabilities. The displayed curve is the mean across seeds, while each reported PR AUC is the mean $\pm$ standard deviation of the ten per-seed PR AUC values and is not calculated from the displayed mean curve.}
    \label{fig:PR_curves}
\end{figure*}

\begin{figure*}[ht!]
    \centering
    \includegraphics[width=\textwidth]{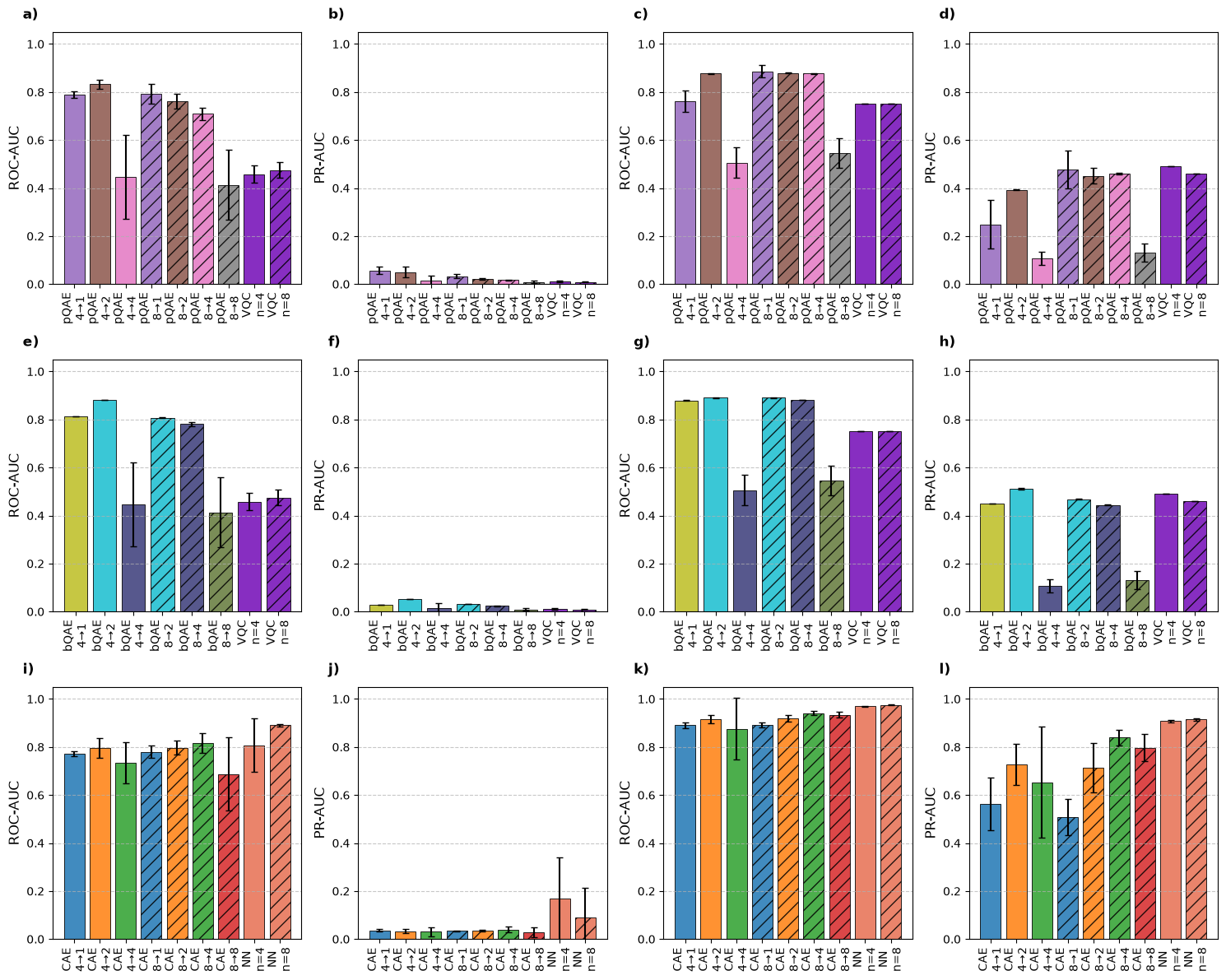}
    \caption{Mean ROC AUC and PR AUC values averaged across ten optimization seeds. Error bars indicate one standard deviation. Higher values correspond to improved discrimination between normal and anomalous samples. Rows compare pQAE with VQC, bQAE with VQC, and CAE with NN. Columns show Kepler ROC AUC, Kepler PR AUC, HTRU-2 ROC AUC, and HTRU-2 PR AUC, respectively. Panels (a)--(l) follow this row-major order.}
    \label{fig:ROC_PR}
\end{figure*}

\begin{figure*}[ht!]
    \centering
    \includegraphics[width=0.80\textwidth]{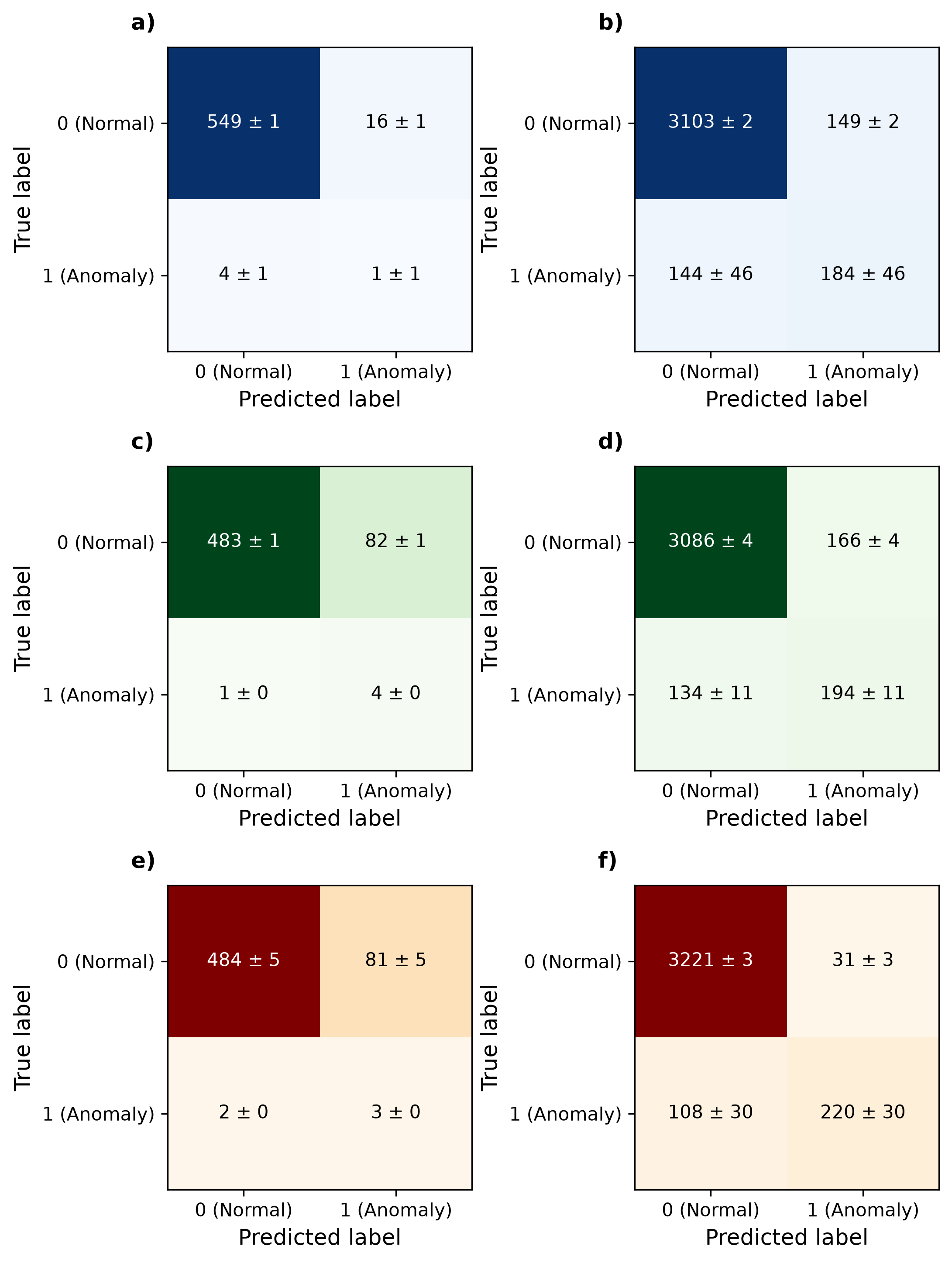}
    \caption{Confusion matrices for the representative pQAE, bQAE, and CAE configurations on Kepler and HTRU-2. Within each autoencoder family and dataset, the configuration with the highest mean ROC AUC was selected. Its displayed reconstruction-error percentile was then selected by the highest mean F1 score across the ten optimization seeds. The rows show pQAE, bQAE, and CAE, while the columns show Kepler and HTRU-2. The selected configurations are pQAE $4\rightarrow2$, bQAE $4\rightarrow2$, and CAE $8\rightarrow4$ for Kepler, and pQAE $8\rightarrow1$, bQAE $4\rightarrow2$, and CAE $8\rightarrow4$ for HTRU-2. Anomalies correspond to label 1, while normal samples correspond to label 0. Matrix entries are the rounded mean count $\pm$ rounded standard deviation across the ten optimization seeds.}
    \label{fig:cm}
\end{figure*}

The ROC curves are shown in Figure~\ref{fig:ROC_curves}. The plotted ROC curves show the mean across ten optimization seeds, with shaded regions indicating one standard deviation.

The corresponding precision--recall curves are shown in Figure~\ref{fig:PR_curves}. ROC curves summarize the trade-off between true-positive and false-positive rates across all possible thresholds, providing a threshold-independent measure of model discrimination. However, ROC curves can be overly optimistic in highly imbalanced settings, making the PR AUC an important complementary metric for evaluating performance on the minority positive class \cite{saito2015precision}. The positive-class prevalence is $5/570=0.0088$ for Kepler and $328/3580=0.0916$ for HTRU-2. ROC AUC, PR AUC, and supplementary average precision were calculated separately for every run from its raw test labels and scores, independently of the plotted mean curves.

The bar graphs in Figure~\ref{fig:ROC_PR} show the average ROC AUC and PR AUC values across ten optimization seeds, with error bars indicating one standard deviation. AUC values closer to 1 indicate stronger separation between normal and anomalous samples.

For the pQAE architectures, the ROC curves in Figures~\ref{fig:ROC_curves}(a, b) indicate different performance across the latent-space sizes, with overlap between several uncertainty regions. For Kepler, the $4\rightarrow2$ pQAE achieves the highest mean values within the family, with ROC AUC $0.833\pm0.019$ and PR AUC $0.050\pm0.023$. For HTRU-2, the $8\rightarrow1$ pQAE is strongest within the family, with ROC AUC $0.887\pm0.026$ and PR AUC $0.476\pm0.079$. The Kepler PR AUC values remain low for all pQAE models, indicating limited ability to reliably identify the five anomalous test samples.

The bQAE architectures exhibit a clearer separation between strong and weak performers in Figures~\ref{fig:ROC_curves}(c, d). For both datasets, the $4\rightarrow2$ bQAE achieves the highest mean values within the family. It reaches ROC AUC $0.882\pm0.001$ and PR AUC $0.052\pm0.000$ on Kepler, and ROC AUC $0.890\pm0.002$ and PR AUC $0.513\pm0.003$ on HTRU-2. The uncertainty regions are generally smaller than those observed for several pQAE configurations, indicating more consistent behaviour across optimization seeds.

An interesting observation is that the full-latent pQAE and bQAE controls are structurally identical at matched input dimensions and produce identical results. Both operate without latent compression, and their ROC AUC values remain close to random discrimination. At every available compressed matched pair $(n,k)$, the bQAE has a higher mean ROC AUC than the pQAE on both datasets. These comparisons use the same input representation and latent dimension, allowing the difference in latent treatment to be examined separately from input dimensionality.

The classical autoencoders exhibit larger uncertainty regions than the quantum autoencoders in some configurations, as shown in Figures~\ref{fig:ROC_curves}(e, f), indicating greater run-to-run variability. For both datasets, the $8\rightarrow4$ CAE achieves the highest mean ROC AUC within the family. It reaches ROC AUC $0.816\pm0.041$ and PR AUC $0.040\pm0.011$ on Kepler, and ROC AUC $0.941\pm0.008$ and PR AUC $0.839\pm0.034$ on HTRU-2.

The NN achieves the highest ROC AUC values of all investigated models on both datasets, as shown in Figure~\ref{fig:ROC_PR}. The eight-component NN reaches ROC AUC $0.891\pm0.006$ and PR AUC $0.091\pm0.122$ on Kepler, and ROC AUC $0.976\pm0.001$ and PR AUC $0.915\pm0.006$ on HTRU-2. The VQC performs weakly on Kepler and moderately on HTRU-2. The eight-component Kepler VQC reaches ROC AUC $0.475\pm0.032$, while the four-component HTRU-2 VQC reaches $0.751\pm0.000$. Both the NN and VQC are trained directly using label information and weighted binary cross-entropy loss, allowing them to explicitly optimize class separation. In contrast, the autoencoder-based approaches are trained only to reconstruct normal samples and therefore rely on reconstruction error as an indirect indicator of anomalies. Ranking the strongest configuration from each family by mean ROC AUC gives NN, bQAE, pQAE, CAE, and VQC on Kepler. On HTRU-2, the order is NN, CAE, bQAE, pQAE, and VQC.

\subsection{Evaluation Metrics and Confusion Matrices}

The reconstruction-based models were compared with the classical and quantum benchmark methods using standard evaluation metrics. Thresholds at the 70th, 80th, 95th, and 99th percentiles of the normal-training reconstruction-error distribution were evaluated. A sample was classified as anomalous if its anomaly score exceeded the selected threshold. Within each autoencoder family and dataset, the representative configuration was selected by its highest mean ROC AUC, and its displayed percentile was selected by the highest mean test F1 score. The resulting confusion matrices are shown in Figure~\ref{fig:cm}, and the corresponding evaluation metrics are reported in Table~\ref{table:ad_metrics}.

\begin{table*}[ht!]
    \centering
    \setlength{\tabcolsep}{4pt}
    \fontsize{8pt}{10pt}\selectfont
    \caption{Performance metrics for the representative pQAE, bQAE, and CAE configuration for each dataset at its selected reconstruction-error percentile. The configuration in each family was selected by the highest mean ROC AUC, and the displayed percentile by the highest mean test F1 score. Values are reported as mean $\pm$ sample standard deviation across ten optimization seeds.}
    \resizebox{\textwidth}{!}{\begin{tabular}{lllcccccccccc}
        \hline\hline
        Dataset & Model & Configuration & Percentile & ROC AUC & PR AUC & AP & Accuracy & Precision & Recall & F1 & Specificity \\
        \hline
        Kepler & pQAE & $4\rightarrow2$ & 95th & $0.833\pm0.019$ & $0.050\pm0.023$ & $0.068\pm0.032$ & $0.965\pm0.003$ & $0.079\pm0.032$ & $0.280\pm0.103$ & $0.123\pm0.048$ & $0.971\pm0.002$ \\
        Kepler & bQAE & $4\rightarrow2$ & 80th & $0.882\pm0.001$ & $0.052\pm0.000$ & $0.066\pm0.001$ & $0.854\pm0.001$ & $0.046\pm0.000$ & $0.800\pm0.000$ & $0.087\pm0.001$ & $0.854\pm0.001$ \\
        Kepler & CAE & $8\rightarrow4$ & 80th & $0.816\pm0.041$ & $0.040\pm0.011$ & $0.053\pm0.014$ & $0.855\pm0.008$ & $0.035\pm0.004$ & $0.580\pm0.063$ & $0.066\pm0.008$ & $0.857\pm0.008$ \\
        HTRU-2 & pQAE & $8\rightarrow1$ & 95th & $0.887\pm0.026$ & $0.476\pm0.079$ & $0.478\pm0.079$ & $0.918\pm0.012$ & $0.542\pm0.075$ & $0.560\pm0.140$ & $0.549\pm0.110$ & $0.954\pm0.001$ \\
        HTRU-2 & bQAE & $4\rightarrow2$ & 95th & $0.890\pm0.002$ & $0.513\pm0.003$ & $0.515\pm0.003$ & $0.916\pm0.002$ & $0.539\pm0.010$ & $0.593\pm0.035$ & $0.564\pm0.022$ & $0.949\pm0.001$ \\
        HTRU-2 & CAE & $8\rightarrow4$ & 99th & $0.941\pm0.008$ & $0.839\pm0.034$ & $0.839\pm0.034$ & $0.961\pm0.008$ & $0.877\pm0.017$ & $0.671\pm0.092$ & $0.757\pm0.066$ & $0.991\pm0.001$ \\
        \hline\hline
    \end{tabular}}
    \label{table:ad_metrics}
\end{table*}

\begin{table*}[t]
    \centering
    \fontsize{8pt}{10pt}\selectfont
    \caption{Performance metrics for the supervised baseline models at both evaluated input dimensions. Classification uses an anomaly-probability cutoff of 0.5. Values are reported as mean $\pm$ sample standard deviation across ten optimization seeds.}
    \resizebox{\textwidth}{!}{\begin{tabular}{lllcccccccc}
        \hline\hline
        Dataset & Model & Input & ROC AUC & PR AUC & AP & Accuracy & Precision & Recall & F1 & Specificity \\
        \hline
        Kepler & VQC & 4 & $0.458\pm0.036$ & $0.010\pm0.003$ & $0.014\pm0.005$ & $0.773\pm0.039$ & $0.015\pm0.004$ & $0.380\pm0.063$ & $0.029\pm0.007$ & $0.776\pm0.040$ \\
        Kepler & VQC & 8 & $0.475\pm0.032$ & $0.009\pm0.001$ & $0.012\pm0.002$ & $0.858\pm0.035$ & $0.010\pm0.008$ & $0.160\pm0.126$ & $0.018\pm0.015$ & $0.864\pm0.036$ \\
        Kepler & NN & 4 & $0.807\pm0.111$ & $0.167\pm0.173$ & $0.037\pm0.014$ & $0.723\pm0.104$ & $0.021\pm0.008$ & $0.740\pm0.299$ & $0.040\pm0.015$ & $0.723\pm0.108$ \\
        Kepler & NN & 8 & $0.891\pm0.006$ & $0.091\pm0.122$ & $0.058\pm0.011$ & $0.719\pm0.071$ & $0.030\pm0.006$ & $0.960\pm0.084$ & $0.059\pm0.011$ & $0.717\pm0.072$ \\
        HTRU-2 & VQC & 4 & $0.751\pm0.000$ & $0.491\pm0.000$ & $0.491\pm0.000$ & $0.687\pm0.005$ & $0.182\pm0.002$ & $0.689\pm0.004$ & $0.288\pm0.003$ & $0.687\pm0.006$ \\
        HTRU-2 & VQC & 8 & $0.750\pm0.000$ & $0.460\pm0.000$ & $0.461\pm0.000$ & $0.716\pm0.003$ & $0.194\pm0.001$ & $0.666\pm0.008$ & $0.300\pm0.001$ & $0.721\pm0.004$ \\
        HTRU-2 & NN & 4 & $0.970\pm0.001$ & $0.908\pm0.006$ & $0.905\pm0.009$ & $0.958\pm0.004$ & $0.718\pm0.028$ & $0.890\pm0.009$ & $0.794\pm0.015$ & $0.965\pm0.005$ \\
        HTRU-2 & NN & 8 & $0.976\pm0.001$ & $0.915\pm0.006$ & $0.913\pm0.008$ & $0.965\pm0.009$ & $0.768\pm0.065$ & $0.900\pm0.012$ & $0.827\pm0.035$ & $0.972\pm0.011$ \\
        \hline\hline
    \end{tabular}}
    \label{table:baseline_metrics}
\end{table*}

Table~\ref{table:baseline_metrics} reports both evaluated input dimensions for each supervised baseline. On Kepler, the eight-component VQC and NN achieve the higher mean ROC AUC within their respective model families. On HTRU-2, the four-component VQC is marginally higher than the eight-component VQC, while the eight-component NN is higher than the four-component NN.

The confusion matrices in Figure~\ref{fig:cm} are consistent with the reported metrics, showing the large number of true negatives resulting from the class imbalance, together with the true-positive detections. False-negative and false-positive classifications are also present. For Kepler, the selected percentiles are 95th for the pQAE $4\rightarrow2$ configuration and 80th for both the bQAE $4\rightarrow2$ and CAE $8\rightarrow4$ configurations. For HTRU-2, the selected percentiles are 95th for the pQAE $8\rightarrow1$ and bQAE $4\rightarrow2$ configurations, and 99th for the CAE $8\rightarrow4$ configuration. These selections carry an inherent optimistic bias because the reported test results were also used to choose the displayed threshold. They also show the trade-off between anomaly recall and false-positive control when continuous reconstruction scores are converted to binary decisions.

Unfortunately, the preferred reconstruction model depends on the dataset and the chosen metric. On Kepler, the selected pQAE obtains the largest anomaly F1 score among the three autoencoders, $0.123\pm0.048$, while the selected bQAE obtains the highest recall, $0.800\pm0.000$. The bQAE's higher recall is accompanied by more false-positive normal samples and lower precision than the selected pQAE. These quantities remain coarse because the five test anomalies make each individual detection correspond to a recall increment of 0.2. The low Kepler precision and PR AUC values show that the false-positive burden remains substantial even when ROC AUC is comparatively high.

Since the focus of this study is anomaly detection, particular attention should be given to metrics that evaluate performance on the positive class, namely precision, recall, and anomaly F1 score. HTRU-2 provides a more informative positive-class evaluation through its 328 test anomalies. Its selected pQAE and bQAE achieve F1 scores of $0.549\pm0.110$ and $0.564\pm0.022$, respectively. The selected CAE achieves precision $0.877\pm0.017$, recall $0.671\pm0.092$, and F1 $0.757\pm0.066$. Although these values indicate that a substantial fraction of anomalies can be detected, they also imply non-negligible false-negative rates, which remain undesirable in anomaly-detection applications. The CAE architecture was intentionally kept relatively simple to provide a meaningful analogue to the quantum models. Classical architectures can generally be improved through the addition of layers, nodes, or more sophisticated training strategies with comparatively little computational overhead, whereas increasing the complexity of quantum models introduces additional runtime costs and, on real devices, greater susceptibility to noise. Consequently, the matched results support an advantage for the explicit quantum bottleneck over the persistent quantum architecture, while the classical autoencoder remains stronger overall on HTRU-2.

The fully supervised benchmark methods are trained directly using class labels. This behaviour is reflected in the ROC curves and AUC values discussed previously. Direct metric comparisons show that the NN achieves the strongest overall anomaly-discrimination performance on both datasets. The VQC performs substantially worse on Kepler, with ROC AUC $0.475\pm0.032$, and shows moderate discrimination on HTRU-2, with ROC AUC $0.751\pm0.000$. The NN can be scaled relatively easily through the addition of hidden nodes or layers, potentially increasing its expressive power further. In contrast, increasing the complexity of the VQC requires additional quantum resources and deeper circuits, which may increase training difficulty and hardware noise. As a result, the practical advantage currently remains with the classical supervised approach.

\subsection{Paired ROC AUC Comparisons}

Table~\ref{table:matched_auc_tests} summarizes the exact paired sign-flip tests and Holm-adjusted results for the structurally matched comparison families. For both datasets, the bQAE has a higher ROC AUC than the pQAE in every compressed matched comparison. All four differences remain significant after correction, with adjusted $p=0.0078$ in each dataset. This provides consistent optimization-seed evidence that the explicit bottleneck improves ROC AUC relative to the persistent architecture at matched input and latent dimensions.

\begin{table*}[t]
    \centering
    \setlength{\tabcolsep}{4pt}
    \renewcommand{\arraystretch}{1.15}
    \fontsize{7.5pt}{9pt}\selectfont
    \caption{Exact two-sided paired sign-flip tests on seed-level ROC AUC differences for the three structurally matched comparison families. A separate Holm correction was applied within each family. ``Significant'' gives the number of comparisons with Holm-adjusted $p<0.05$. The final column reports the adjusted $p$-value or range for the significant comparisons.}
    \begin{tabularx}{\textwidth}{@{}llcc>{\raggedright\arraybackslash}Xl@{}}
        \hline\hline
        Dataset & Family & Tests & Significant & Significant differences & Holm-adjusted $p$ \\
        \hline
        Kepler & bQAE vs. pQAE & 4 & 4 & bQAE higher in 4/4 & 0.0078 \\
        Kepler & bQAE vs. CAE & 6 & 4 & bQAE higher at $4\rightarrow1$ and $4\rightarrow2$; CAE higher at $4\rightarrow4$ and $8\rightarrow8$ & 0.0117--0.0234 \\
        Kepler & pQAE vs. CAE & 7 & 3 & CAE higher at $4\rightarrow4$, $8\rightarrow4$, and $8\rightarrow8$ & 0.0137--0.0352 \\
        HTRU-2 & bQAE vs. pQAE & 4 & 4 & bQAE higher in 4/4 & 0.0078 \\
        HTRU-2 & bQAE vs. CAE & 6 & 6 & CAE higher in 6/6 & 0.0117 \\
        HTRU-2 & pQAE vs. CAE & 7 & 6 & CAE higher in 6/7; $8\rightarrow1$ is inconclusive & 0.0137 \\
        \hline\hline
    \end{tabularx}
    \label{table:matched_auc_tests}
\end{table*}

\begin{table*}[t]
    \centering
    \setlength{\tabcolsep}{4pt}
    \renewcommand{\arraystretch}{1.15}
    \fontsize{7.5pt}{9pt}\selectfont
    \caption{Exact two-sided paired sign-flip tests on seed-level ROC AUC differences for the data-selected autoencoders against input-matched supervised benchmarks. Positive $\Delta$ROC AUC favours the selected autoencoder. ``Wins'' gives the number of the ten paired optimization seeds for which the selected autoencoder has a higher ROC AUC than the listed comparator. The same ten optimization runs were used for model selection and statistical testing. The Holm adjustment addresses multiplicity within the reported six-comparison family but does not account for the preceding architecture-selection step.}
    \begin{tabular}{@{}l>{\raggedright\arraybackslash}p{0.24\textwidth}cccc@{}}
        \hline\hline
        Dataset & Selected comparison & $\Delta$ROC AUC & Wins & Exact $p$ & Holm-adjusted $p$ \\
        \hline
        Kepler & pQAE $4\rightarrow2$ vs. VQC$_4$ & $+0.375$ & 10/10 & 0.0020 & 0.0117 \\
        Kepler & pQAE $4\rightarrow2$ vs. NN$_4$ & $+0.025$ & 4/10 & 0.8203 & 0.8203 \\
        Kepler & bQAE $4\rightarrow2$ vs. VQC$_4$ & $+0.424$ & 10/10 & 0.0020 & 0.0117 \\
        Kepler & bQAE $4\rightarrow2$ vs. NN$_4$ & $+0.074$ & 10/10 & 0.0020 & 0.0117 \\
        Kepler & CAE $8\rightarrow4$ vs. VQC$_8$ & $+0.341$ & 10/10 & 0.0020 & 0.0117 \\
        Kepler & CAE $8\rightarrow4$ vs. NN$_8$ & $-0.075$ & 0/10 & 0.0020 & 0.0117 \\
        HTRU-2 & pQAE $8\rightarrow1$ vs. VQC$_8$ & $+0.136$ & 10/10 & 0.0020 & 0.0117 \\
        HTRU-2 & pQAE $8\rightarrow1$ vs. NN$_8$ & $-0.089$ & 0/10 & 0.0020 & 0.0117 \\
        HTRU-2 & bQAE $4\rightarrow2$ vs. VQC$_4$ & $+0.139$ & 10/10 & 0.0020 & 0.0117 \\
        HTRU-2 & bQAE $4\rightarrow2$ vs. NN$_4$ & $-0.079$ & 0/10 & 0.0020 & 0.0117 \\
        HTRU-2 & CAE $8\rightarrow4$ vs. VQC$_8$ & $+0.191$ & 10/10 & 0.0020 & 0.0117 \\
        HTRU-2 & CAE $8\rightarrow4$ vs. NN$_8$ & $-0.034$ & 0/10 & 0.0020 & 0.0117 \\
        \hline\hline
    \end{tabular}
    \label{table:selected_auc_tests}
\end{table*}

The matched comparisons with the CAE show a stronger dataset dependence. On Kepler, four of the six bQAE--CAE comparisons remain significant after Holm adjustment. The bQAE is higher at $4\rightarrow1$ and $4\rightarrow2$, while the CAE is higher for the two no-compression controls. The matched $8\rightarrow4$ and $8\rightarrow2$ differences are inconclusive. Three of seven Kepler pQAE--CAE comparisons are significant, and all three favour the CAE. On HTRU-2, all six bQAE--CAE comparisons favour the CAE after correction. Six of seven pQAE--CAE comparisons also favour the CAE, with only the $8\rightarrow1$ comparison remaining inconclusive.

The data-selected comparisons in Table~\ref{table:selected_auc_tests} provide a direct connection to the supervised benchmarks. On Kepler, each selected autoencoder exceeds the input-matched VQC after correction. The selected bQAE also exceeds the four-input NN, the selected pQAE--NN comparison is inconclusive after correction, and the eight-input NN exceeds the selected CAE. On HTRU-2, all three selected autoencoders exceed their input-matched VQC and are exceeded by their input-matched NN. These results describe optimization-seed consistency for the selected models. The model-selection step and fixed test partitions limit broader statistical generalization.

\subsection{Limitations}

Using the same fixed partition for every model allows direct comparisons on the same test samples. The seed bands therefore show how much the results change with optimization, while variation from alternative data partitions remains outside their scope. This is particularly important for Kepler because its five test anomalies make recall change in increments of 0.2 and cause PR estimates to depend strongly on individual predictions. HTRU-2 has a much larger positive class, although it is also evaluated using one fixed partition. The Kepler feature-selection step uses the supplied training labels before the autoencoders are trained on normal samples, so the resulting pipeline is semi-supervised and includes label-informed preprocessing.

The paired tests inherit the same scope. Their $p$-values measure the consistency of within-seed ROC AUC differences on fixed test samples and do not include uncertainty from alternative partitions or test-sample resampling.

 The representative configurations and displayed percentiles in Table~\ref{table:ad_metrics} and Figure~\ref{fig:cm} were selected from the observed results and therefore provide optimistic summaries of the evaluated runs. Formal pairwise testing focused on comparisons between architecture families, so the configurations with the highest observed mean ROC AUC should be interpreted as observed within-family rankings. The CAE was deliberately kept close to the quantum architectures, so a more extensively tuned classical model could change the comparison. Clipping also removes the magnitude of values beyond the fitted angle range. The Kepler clipping count was zero, while one of the 328 HTRU-2 test anomalies contained a clipped component. Finally, the quantum experiments use analytic noiseless simulation. The 15-qubit bQAE $8\rightarrow1$ configuration was impractical to complete within the runtime and memory available on the local hardware used for this study, leaving that matched comparison incomplete.

\section{Conclusion}\label{sec:conc}

The results demonstrate that quantum convolutional autoencoders can be applied to reconstruction-based anomaly detection on the Kepler exoplanet and HTRU-2 pulsar-candidate datasets. Overall anomaly-detection performance varied between the two datasets, particularly for precision, recall, and anomaly F1 score. Among the quantum architectures, the bQAE achieved a higher mean ROC AUC than the pQAE at every available compressed matched input and latent dimension. All four compressed matched differences remain significant after Holm adjustment on both datasets, supporting the consistency of the bottleneck advantage across optimization seeds. The highest observed mean ROC AUC values occurred in compressed configurations, suggesting that some degree of latent-space compression may be beneficial. The preferred compressed latent-space size depends on the dataset and model family.

Comparison with the classical autoencoder showed that the best bQAE mean ROC AUC exceeds the best CAE mean on Kepler, while the CAE achieves stronger ROC AUC, PR AUC, and thresholded F1 performance on HTRU-2 under the architectural constraints considered here. The paired tests reinforce this dataset dependence: the significant Kepler bQAE--CAE differences favour different families across configurations, while every matched HTRU-2 bQAE--CAE comparison significantly favours the CAE after correction. These findings support a dataset-dependent comparison rather than a general quantum advantage. The CAE was intentionally designed as a simple analogue of the quantum architectures rather than as a highly optimized classical model. Classical architectures can be expanded and refined through additional layers, nodes, and training improvements with comparatively little computational overhead, whereas greater quantum-model complexity introduces additional runtime costs and, on quantum hardware, greater susceptibility to noise. The CAE therefore remains the stronger overall reconstruction model on HTRU-2. The Kepler comparison remains limited by its five anomalous test samples and should be interpreted within the model and simulator constraints considered here.

The NN achieves the strongest continuous-score discrimination on both datasets. The VQC produces substantially weaker results, with below-random mean ROC AUC on Kepler and moderate discrimination on HTRU-2. The data-selected paired comparisons place all three selected autoencoders above the input-matched VQC on both datasets after correction. On HTRU-2, the input-matched NN significantly exceeds every selected autoencoder. The corresponding Kepler comparisons are mixed. Since both supervised models are trained directly using class labels, they explicitly optimize class separation. The NN can also be scaled relatively easily through additional hidden nodes or layers. Increasing the complexity of the VQC instead requires additional quantum resources and deeper circuits, which may increase training difficulty and hardware noise. The practical advantage therefore remains with the classical supervised approach.

Future work could explore alternative convolutional and pooling structures, different entangling operations, and a wider range of latent-space sizes. Extending the analysis to additional anomaly-detection datasets and real quantum hardware would provide further insight into the generalizability and practical limitations of these architectures.
\section*{Funding}

This work was funded by the South African Quantum Technology Initiative (SA QuTI) through the Department of Science and Innovation of South Africa.

\section*{Data Availability Statement}

 The \href{https://www.kaggle.com/datasets/keplersmachines/kepler-labelled-time-series-data}{Kepler labelled time-series dataset} and the \href{https://archive.ics.uci.edu/ml/datasets/HTRU2}{HTRU-2 dataset} are publicly available for download.

\section*{Conflict of Interest}
The authors have no conflicts of interest to disclose.

\section*{AI Declaration}

During the preparation and revision of this manuscript, the authors used OpenAI's Codex and ChatGPT tools, including the GPT-5.6 Sol and GPT-6 Astra models, to assist with drafting and editing text, developing and reviewing code, and conducting internal consistency checks and audits of the reported results. These tools were used particularly to assist with the implementation, checking, and presentation of the Holm multiple-comparison corrections. The research questions, methodological design, model architectures, comparison families, and scientific interpretation were determined by the authors. All AI-assisted material was critically reviewed and approved by the authors, who take full responsibility for the accuracy, integrity, and conclusions of the work.

\bibliography{sn-bibliography}

\end{document}